\documentclass[a4paper,14pt]{article}
\pdfoutput=1
\usepackage{wrapfig} 

\usepackage{extsizes}
\usepackage[T2A]{fontenc}
\usepackage[utf8]{inputenc}
\usepackage[dvips]{graphicx}
\usepackage{cite}

\usepackage{braket}
\usepackage{graphicx} 
\usepackage{amssymb,amsfonts,amsmath,amsthm} 
\usepackage{indentfirst} 
\usepackage[usenames,dvipsnames]{color} 
\usepackage{multirow} 

\usepackage{geometry}
\usepackage{multicol}

\begin{document}

\begin{center}

{\Large{Artificial Quantum Neural Network: quantum neurons, logical elements and tests of convolutional nets. }}\\[9pt]

{\large  V. I. Dorozhinsky$^{1 a}$, O. V. Pavlovsky $^{1,2,3 b}$}\\[6pt]

\parbox{.96\textwidth}{\centering\small\it
$^1$ Faculty of Physics,
Moscow State University,\\
Moscow, 119991, Russia\\
$^2$Institute for Theoretical and Experimental Physics,\\
Moscow, 117218, Russia\\
$^3$ National Research Nuclear University MEPhI, \\
 Moscow, 115409 Russia \\
E-mail: $^a$dorrozhin@gmail.com  $^b$ovp@goa.bog.msu.ru}\\[1cc]
\end{center}

{\parindent5mm

\begin{center}
\textbf{Abstract}
\end{center}
\par
We consider a model of an artificial neural network that uses
quantum-mechanical particles in a two-humped potential as a
neuron. To simulate such a quantum-mechanical system the
Monte-Carlo integration method is used. A form of the
self-potential of a particle and two potentials (exciting and
inhibiting) interaction are proposed. The possibility of
implementing the simplest logical elements, (such as AND, OR and
NOT) based on introduced quantum particles is shown. Further we
show implementation of a simplest convolutional network. Finally
we construct a network that recognizes handwritten symbols, which
shows that in the case of simple architectures, it is possible to
transfer weights from a classical network to a quantum one.
\thispagestyle{empty}
\newpage

\section{Introduction}

The modern development of various fields of science and technology
strongly depends  on progress in computer science.  Such progress
can be associated with the implementation of new technologies of
computing systems as well as with new computational algorithms.
The two main  tendencies in  the development of the computer
science are quantum computing and artificial intelligence.

The logical elements of computer processors become smaller and
smaller with time.  Today the technology  gives us the possibility
to produce such elements very close to quantum limit where quantum
fluctuations became more and more essential.  On the one hand, the
quantum nature of  such elements  can lead to the crisis of the
classical computing technique, but on the another hand  the
quantum properties  of the elements  can be used for realization
of quantum  calculation algorithms.

Another main tendency in modern development of computer science is
the artificial intelligence.  Artificial neural networks  are the
computing systems inspired by the biological neural networks and
consist of the set of connected  calculation nodes (artificial
neurons).  Like in the brain the connections between such nodes
transmit  the  signals from one node to another.  The parameters
of such connections depend on weights.  By changing of these
weights one can modify the transport of the signals in the neural
network. The learning procedure of the neural network just
consists in the tuning of these  weights.  There are many
computing problems can be solved by using of artificial neural
networks like images \cite{lecun} and speech \cite{speech1}
recognitions, machine translation \cite{google} and so on.

Is it possible to combine these two concepts and   propose quantum
system which would work as a neural network? How small such system
could be? In our work we try to find the answer on these
questions.

The integrate-and-fire  neuron is one of the simplest model of
neuron.  In framework of this simplification the neuron's function
is to integrate of external activity of another neurons and
produces a spike if this integrated external activity reaches some
threshold value. In such cases the neuron excitation is
transmitted to other  neurons in the network. The quantum objects
are stochastic by nature and we can say only about probability of
excitation of our quantum neurons. However, the quantum nature of
our neural network will not prevent us from implementation of all
 basic functions of the neural networks. Moreover, the analogy
with quantum computers allows us to hope that this quantum nature
will significantly improve the functionality of such a network.

The main function of the neuron (classical and quantum) should be
the production of spikes or bursts of activity.  In our model we
propose a soliton solutions of a quantum system  as an analog of
such spike. The example the soliton is a well-known kink solution
(or instantons) \cite{polyakov} which is connected with
spontaneous tunnelling processes between vacua in the model of the
quantum particle in the two-humped potential. These processes are
associated with bursts of action which can be used as an analog of
spikes. If we organize connections between quantum neurons so that
a kink in one of the quantum neurons would produced  a kink in
another one then the activity that has arisen in one part of the
network will spread to other parts of our network and our quantum
neural network  will start to work.

Our work is organized as follows. First, we discuss the general
principles of constructing a quantum neural network. As an example
of the simplest implementation, basic logic elements are
considered. Next, we consider a more complicated problem of
recognizing the vertical line. At last part, we show how a quantum
network can be applied to digital numbers recognition  problem and
some principles of quantum neural network learning  are also
discussed.

\section{Formulation of problem}

\subsection{Quantum neurons (Q-neuron)}

\begin{figure}
\centering
\includegraphics[width=0.5\linewidth]{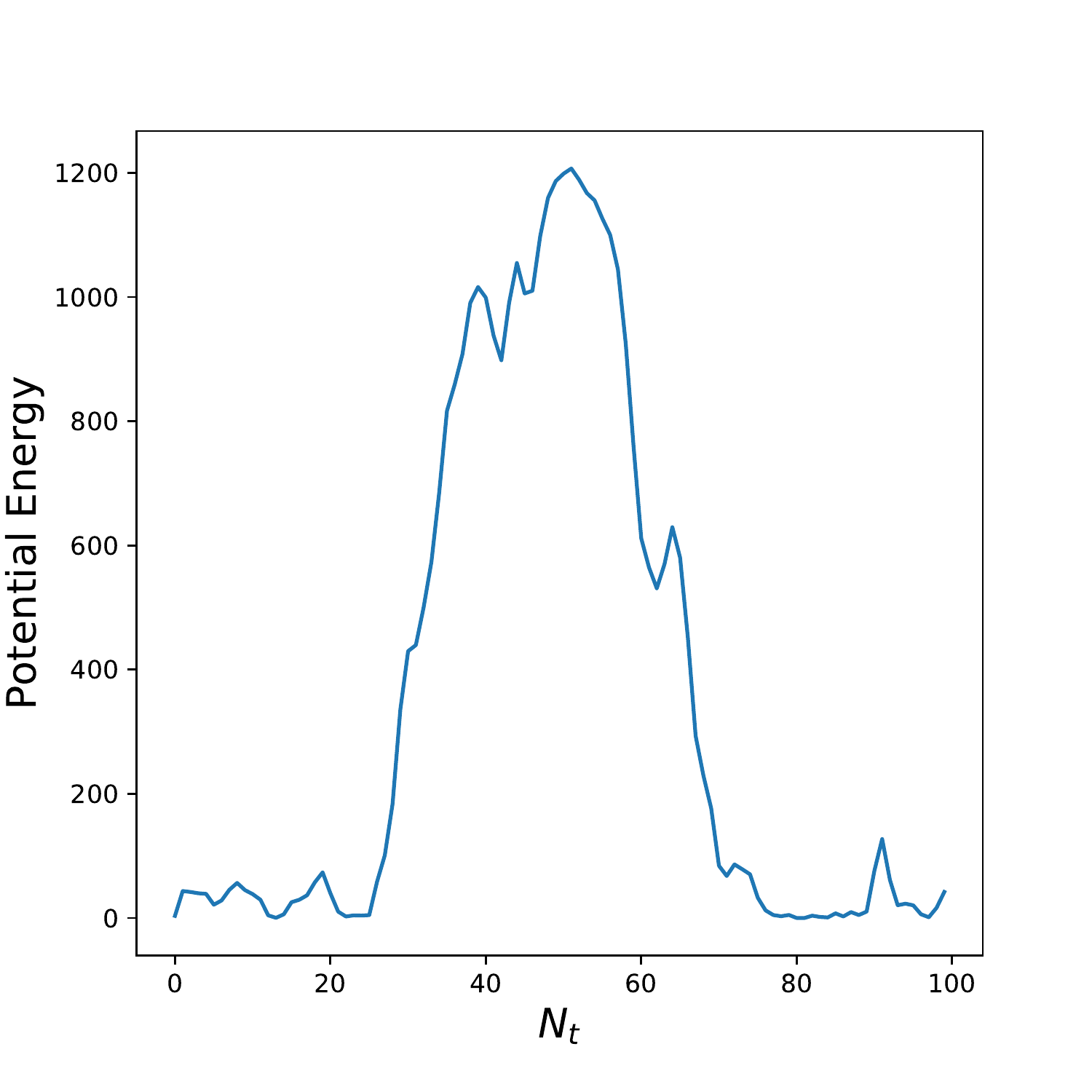}
\caption{Instanton action density. $N_t$ is the index of a time
slice.} \label{fig:instanton}
\end{figure}
Consider a model of a quantum mechanical system organized as a
neural network. A neural network consists of nodes (neurons) and
connections between them (axons). The role of neurons in the model
will be played by quantum-mechanical particles $\hat{p}_{i}$,
evolving under the influence of potential
\begin{equation}
\hat{H}_{i} = \frac{1}{2} \hat{p}_{i}^{2} + V_{0} \left( \hat{\varphi}_{i} \right).
\end{equation}
The role of connections between neurons will be played by the
interaction potential:
\begin{equation}
V_{int} = V_{int} \left( \hat{\varphi}_{i}, \hat{\varphi}_{j} \right).
\end{equation}
Thus, the total Hamiltonian of the system will be following
\begin{equation}
\hat{H} = \sum_{i}\left( \frac{1}{2} \hat{p}_{i}^{2} + V_{0}( \hat{\varphi}_{i} )\right)
+ \sum_{i > j} V_{int}( \hat{\varphi}_{i}, \hat{\varphi}_{j}).\\
\end{equation}
Since in the general case we need to deal with sufficiently
complex quantum-mechanical systems, to describe their properties
we will be using the well-known path integral Monte Carlo
formalism. In Euclidean time, the statistical sum of the system
has the form
\begin{equation}
Z = \int \prod_{i} \mathcal{D} \varphi_{i}\left( \tau \right) \exp(-S(\varphi_{i}(\tau))), \varphi_{i}(0) = \varphi_{i}(T),
\end{equation}
where $\varphi_{i}(\tau)$ -- is the Euclidean path of i-th particle, $\tau \in \left[ 0, T \right] $
-- Euclidean time, and $S(\varphi_{i})$ -- classical action:
\begin{equation}
S = \int_{0}^{T}d\tau \left[ \sum_{i}\left(  \frac{1}{2} \dot{\varphi}_{i}^{2} + V_{0}( \varphi_{i} ) \right)
+ \sum_{i > j} V_{int}( \varphi_{i}, \varphi_{j}) \right].
\end{equation}\par
The observables in such formalism are calculated as
\begin{equation}
\braket{\mathcal{O}(\varphi_{1},...,\varphi_{i})} =
\frac{1}{Z} \int \prod_{i} \mathcal{D}\varphi_{i}(\tau)\mathcal{O}(\varphi_{1},...,\varphi_{i})\exp(-S(\varphi_{i})).
\end{equation}
We now choose the intrinsic potential of the neuron
$V_{0}(\varphi_{i})$. The main function of a neuron is to generate
spikes. Suitable feature can be provided by a particle in the
W-potential:
\begin{equation}
V_0(\varphi_i) = \frac{\Lambda}{4}\left( \varphi^2 - 1 \right)^2.
\end{equation}
A typical quantum-mechanical behavior of such a particle is a
vacuum fluctuation and a rapid change of the vacuum state
(instanton). The instantons are accompanied by a peak of the
action density (Fig.~\ref{fig:instanton}), similar to the
potential of the spike of a biological neuron.

The explicit form of interaction between neurons will be
introduced in the next Section.

\subsection{Methods of solution}

The operation of the network is based on the propagation of
activity from the input nodes (sensors) to the output ones. As
already noted, the network can consist of many nodes. To study
such a complex quantum system, it is natural to apply the Monte
Carlo method\cite{ceperley}. The idea of the method is to use the
Markov process, to generate paths of particles $\varphi_{i}$ with
a statistical weight proportional to $\exp(-S(\varphi_{i}))$.\par
In our work we use the multilevel algorithm of Metropolis. The
introduction of several levels of the algorithm is caused by the
need to suppress autocorrelations and makes it possible to improve
the computational efficiency.

\section{ Excitatory and inhibitory connections of neurons and logical elements}

\subsection{Single neuron}

\begin{figure}
\centering
\includegraphics[width=0.5\linewidth]{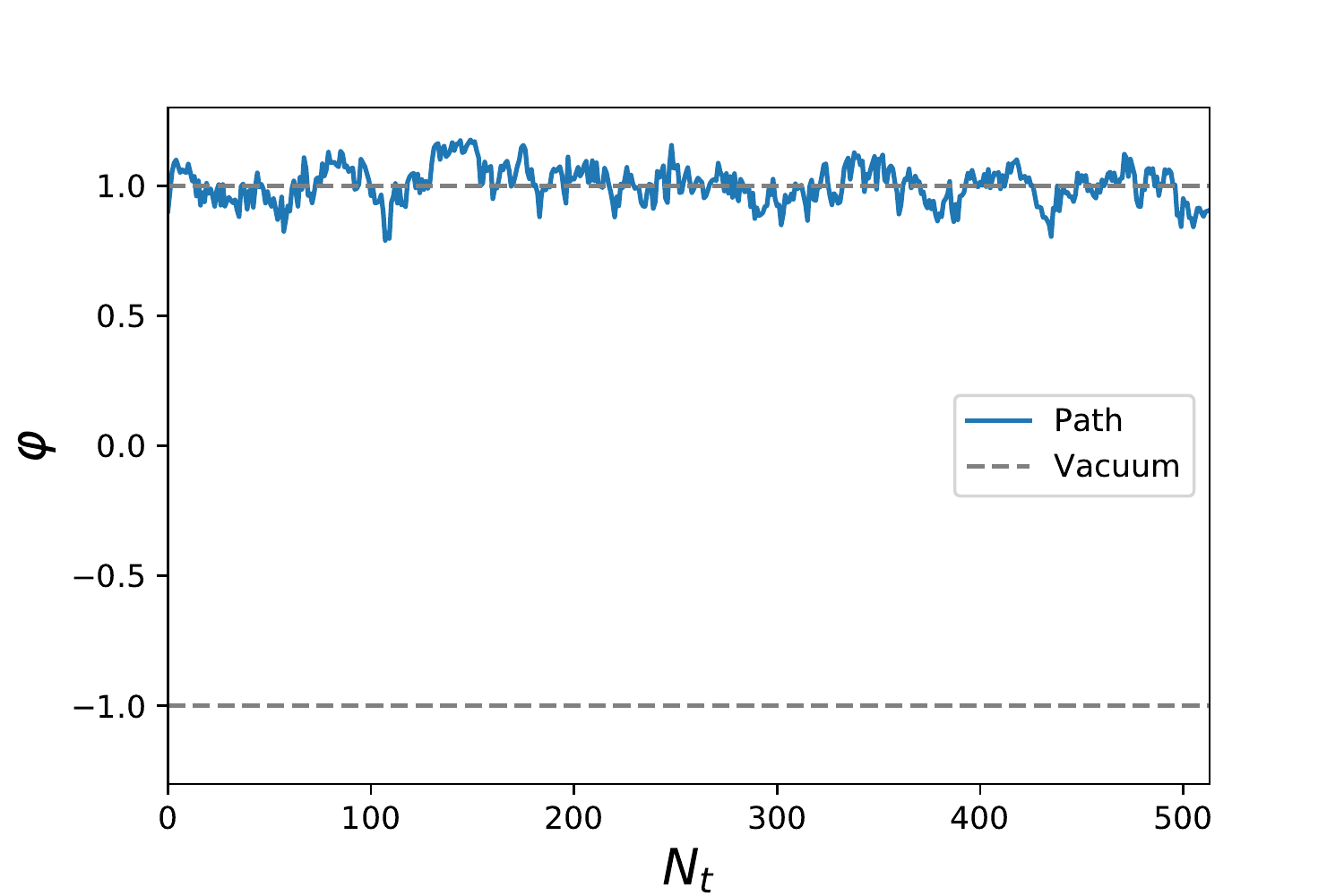}
\caption{Free neuron path. Neuron does not exhibit a spike (i.e.
does not move from one vacuum to another) unless excited.}
\label{fig:singleNeuronPath}
\end{figure}
We start the consideration of our model with a case of a single
neuron. As we have already said, an artificial neuron in our case
will be represented by a particle located in a two-humped
potential. A spike or a single act of activity of such a neuron is
the transition of a particle from one vacuum to another. The
Lagrangian of such a system is written as:
\begin{equation}
\mathcal{L}_{0} = \frac{1}{2}\dot{\varphi}^{2} + \frac{\Lambda}{4} \left(\varphi^2 - 1 \right)^{2}.
\end{equation}
As can be seen from the Lagrangian, the minima of the potential
energy of such a particle are at the points $1$ and $-1$. The
value of the classical action for the transition from one minimum
to another is \cite{polyakov}
\begin{equation}
S_{cl} = \frac{2\sqrt{2\Lambda}}{3}. \label{kink_act}
\end{equation}
This value will be necessary in order to select suitable
$\Lambda$. First, we want the fluctuations of our particle around
the energy minima to be small. Secondly, we want to minimize the
number of spontaneous transitions from one state to another. In
our case, the optimal parameters appeared to be: $\Lambda = 5000$.
Parameters of the Metropolis algorithm were chosen as follows:
time (inverse temperature) $T = 0.7$, number of time grid nodes
$N_t = 512$. To suppress autocorrelations, we use the
thermalization length in $ 2\cdot10^6$ iterations. Finally, note
that for the initialization of the path we use the saw path of 0's
and 1's: $\varphi_{init}=i \bmod 2$ where $i \in [0, 512] \cap
\mathbb{Z}$ is the number of corresponding time grid node. A
typical thermalized path is presented in the
Fig.~\ref{fig:singleNeuronPath}.

\subsection{Two neurons}

We now turn to the case of two interacting neurons. We want a
spike in one of them to cause a spike in the other, but spikes in
the second one should not affect the first one. This means that
the interaction Lagrangian of such neurons must be asymmetric (we
call such a Lagrangian excitatory):
\begin{equation}
\mathcal{L}_{int} =
\varepsilon_{exc}\varphi_{1}^{2}\left(\varphi_{2}^2 - 1\right)^2,
\end{equation}
where $\varepsilon_{exc}$ is the connection strength.    If
$\varphi_{1}$ is in the vacuum, then there is no impact on
$\varphi_2$. But, if $\varphi_{1}$ experiences a spike, then
$\varphi_{2}$ also tends to have a spike. Thus the activity is
spreading from one node to another.

\begin{figure}
    \centering
    \begin{minipage}{0.39\textwidth}
        \centering
        \includegraphics[width=\linewidth]{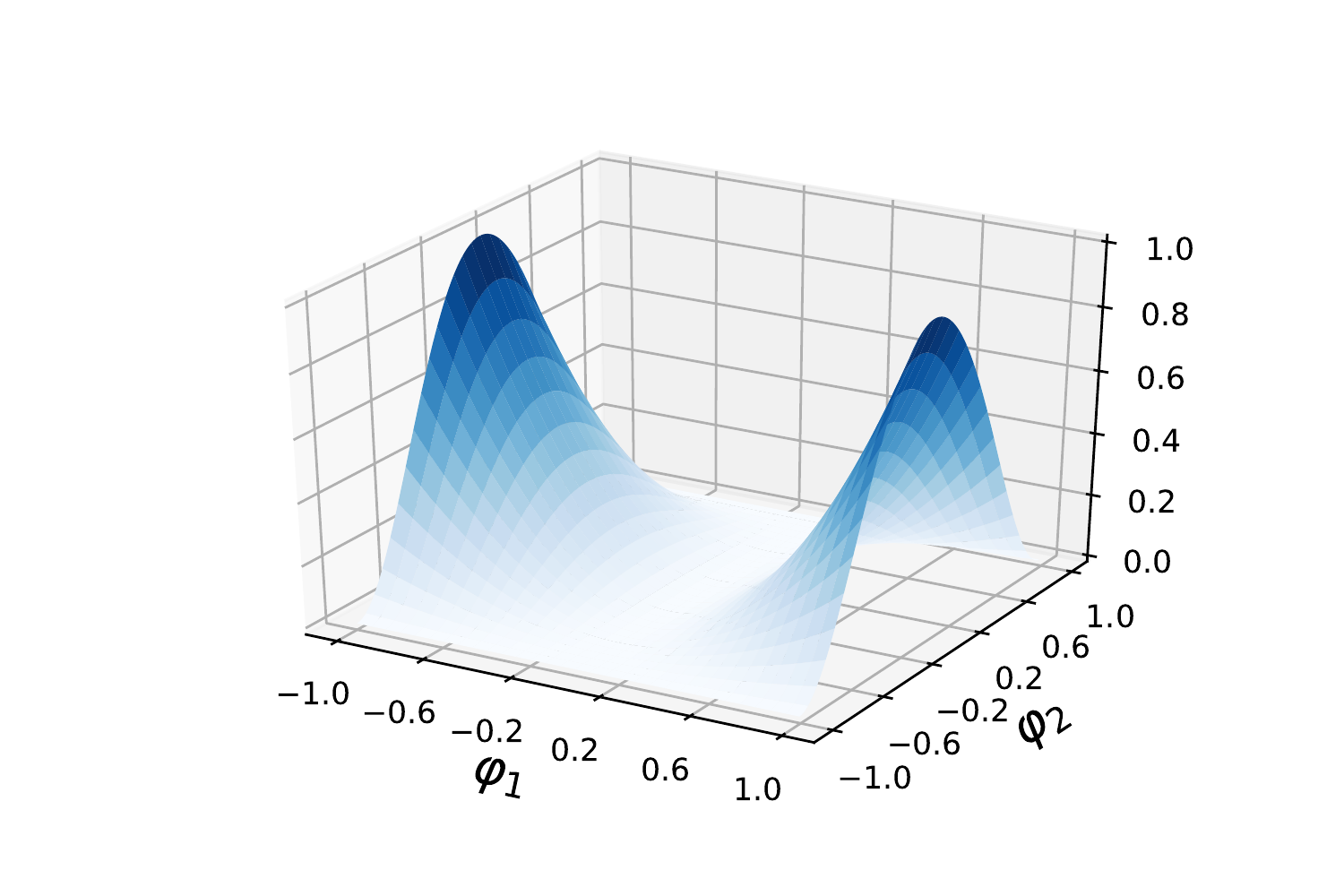}
        \caption{Interaction potential for transmission of spikes from neuron $\varphi_2$ to $\varphi_1$.}
        \label{fig:simpleExcitatory}
    \end{minipage}\hfill
    \begin{minipage}{0.60\textwidth}
        \centering
        \includegraphics[width=\linewidth]{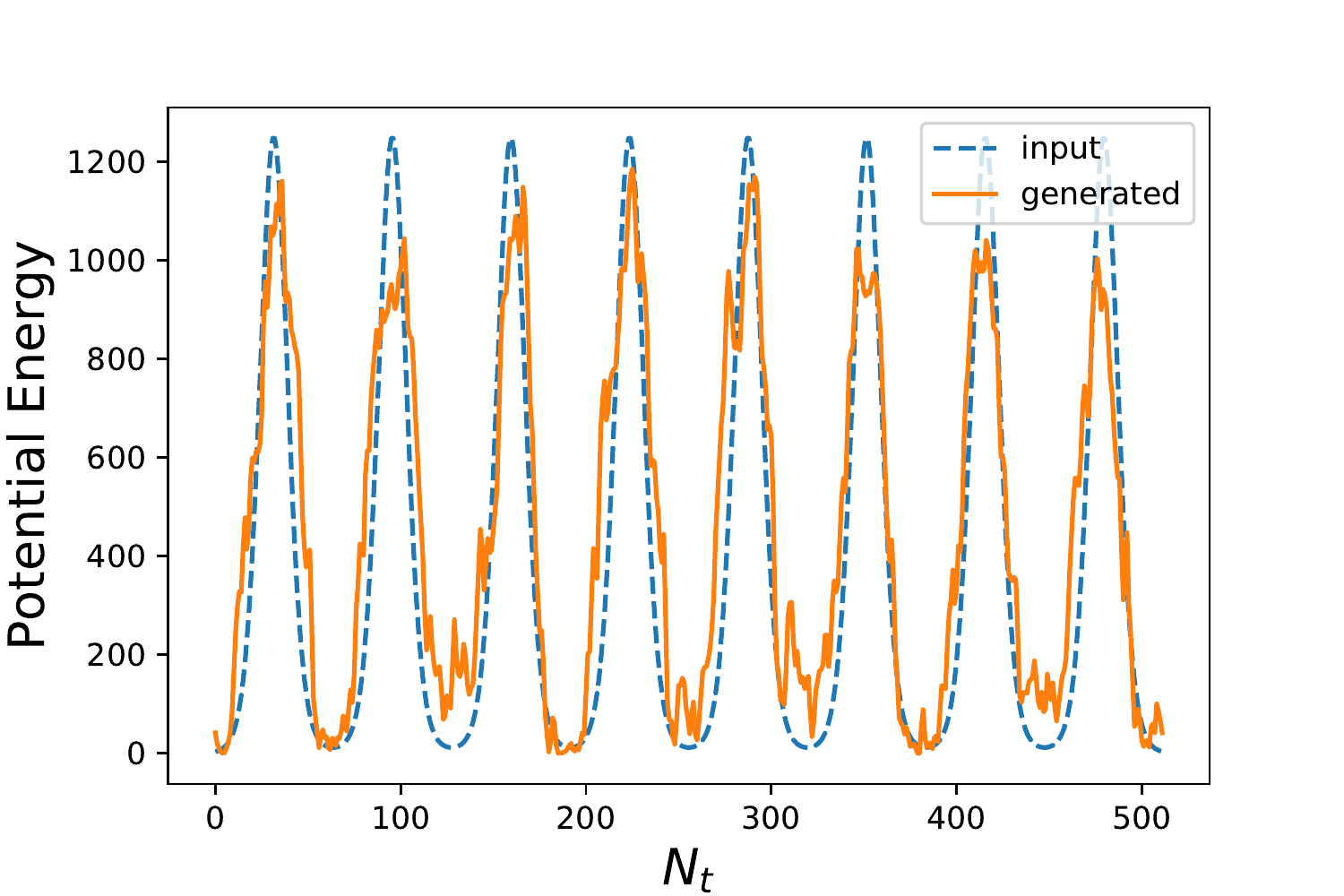} 
        \caption{Potential energy of affected neuron (solid) tend to resemble one of input neuron (dashed).}
        \label{fig:twoNeuronsConnected}
    \end{minipage}
\end{figure}
The plot of the potential of such interaction is shown in
Fig.~\ref{fig:simpleExcitatory}. Axis $\varphi_2$ is related to
the excitatory neuron and $\varphi_1$ corresponds to affected one.
\par

For representing input information, we are going to use input
neurons. Each input neuron can be either passive (make no affect
at all so may be discarded) or be active. Active input neurons
have fixed path (unlike simulated neurons which path evolves
during simulation) which consists of  classical kink solutions.
Potential energy of such an input neuron is depicted in
Fig.~\ref{fig:twoNeuronsConnected} (dashed line). Each peak of
potential energy corresponds to the kink. Solid line represents
the potential energy of the single  neuron affected by the input
neuron.\par

We introduce activity of any simulated neuron as ratio of integral
potential energy of that neuron to input one (e.g. activity of
neuron which never leave vacuum will be 0 and activity of a neuron
which path replicates path of input neuron will be 1). In order to
investigate different schemes we will inspect plots of activity at
some neuron as function of connection strengths
$\varepsilon_{exc}$.\par
\begin{figure}
\centering
\includegraphics[width=0.5\linewidth]{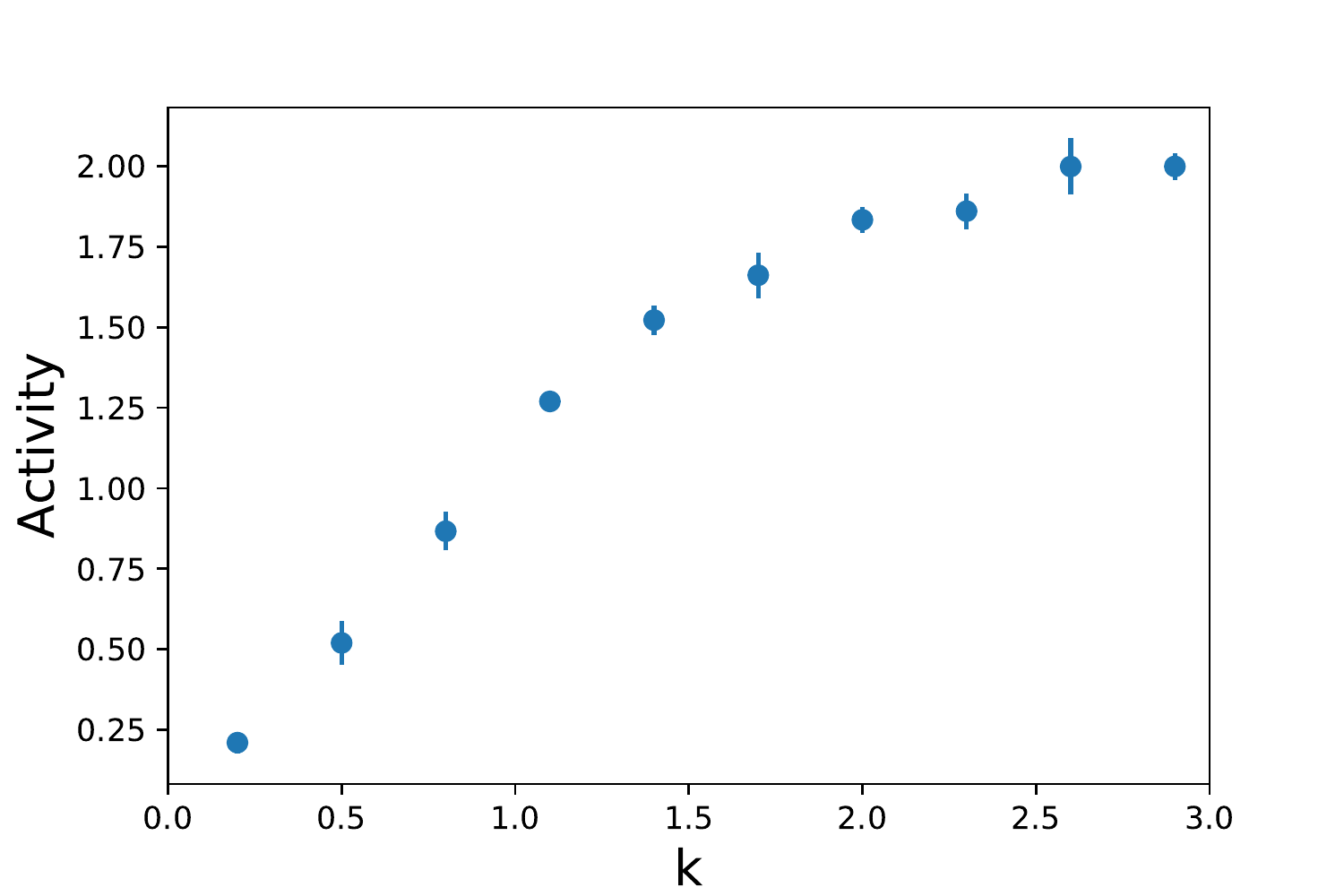}
\caption{
Activity of output neuron grows as function of
connection strength.} \label{fig:SIMPLE}
\end{figure}
In order to study different configurations of neurons we introduce
modulating factor $k$. Once we choose appropriate parameter $\hat
\varepsilon$ for every connection, we multiply each of them by
this factor to obtain new connection strengths $\varepsilon = k
\cdot \hat \varepsilon$ and than plot activity of neuron of
interest as a function of single parameter $k$.\par It was found
that $\varepsilon_{exc}$ can take it's values in the range from
$3000$ to $8000$ (Fig.~\ref{fig:SIMPLE}). In the case of too small
$\varepsilon_{exc}$ neurons almost do not interact and if
$\varepsilon_{exc}$ is too large, their own potentials become
insignificant in comparison with the interaction, which leads to
an undesirable delay of the neuron in the state of $\varphi = 0$.
Plot for $\varepsilon_{exc} = 6000$ is presented in
Fig.~\ref{fig:twoNeuronsConnected}.\par

In the simulation presented in Fig.~\ref{fig:twoNeuronsConnected}
the activity of output neuron appeared to be 0.92.\par
\begin{figure}
\centering
\includegraphics[width=0.5\linewidth]{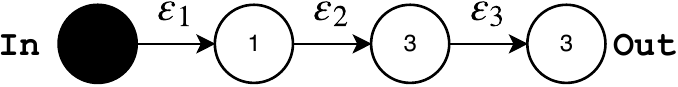}
\caption{Input neuron excites 3 simulated neurons placed in a
row.} \label{fig:LineOf3}
\end{figure}

\begin{figure}
\centering
\includegraphics[width=0.5\linewidth]{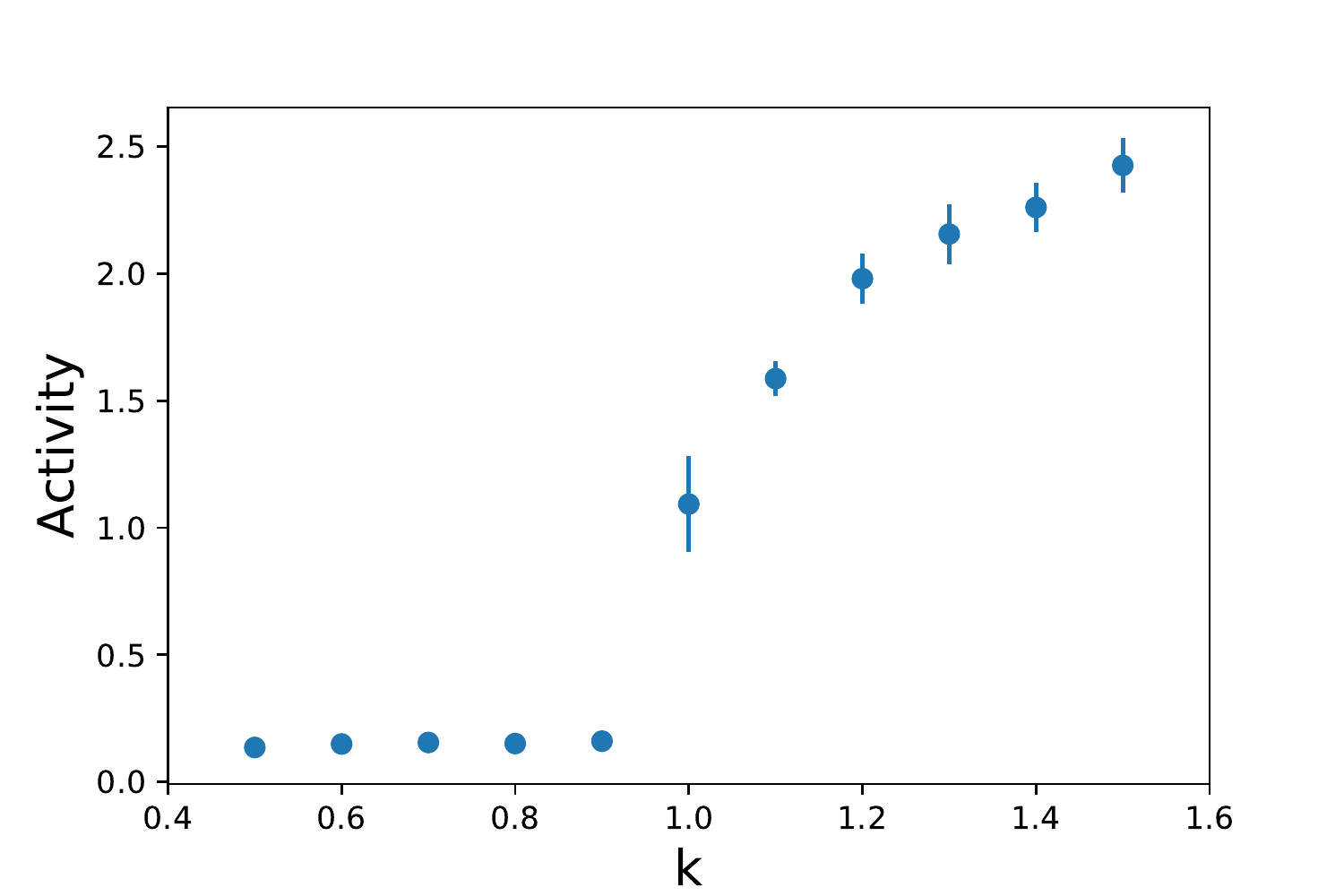}
\caption{ Activity of output neuron (neuron 3) as a function on
connection strength. $\varepsilon_1 = k \cdot 1.5 \cdot 10^4,
\varepsilon_2 = k \cdot 1.0 \cdot 10^4, \varepsilon_3 = k \cdot
0.5 \cdot 10^4$. (see Fig.~\ref{fig:LineOf3})} \label{fig:LINE_3}
\end{figure}

\begin{figure}
\centering
\includegraphics[width=0.7\linewidth]{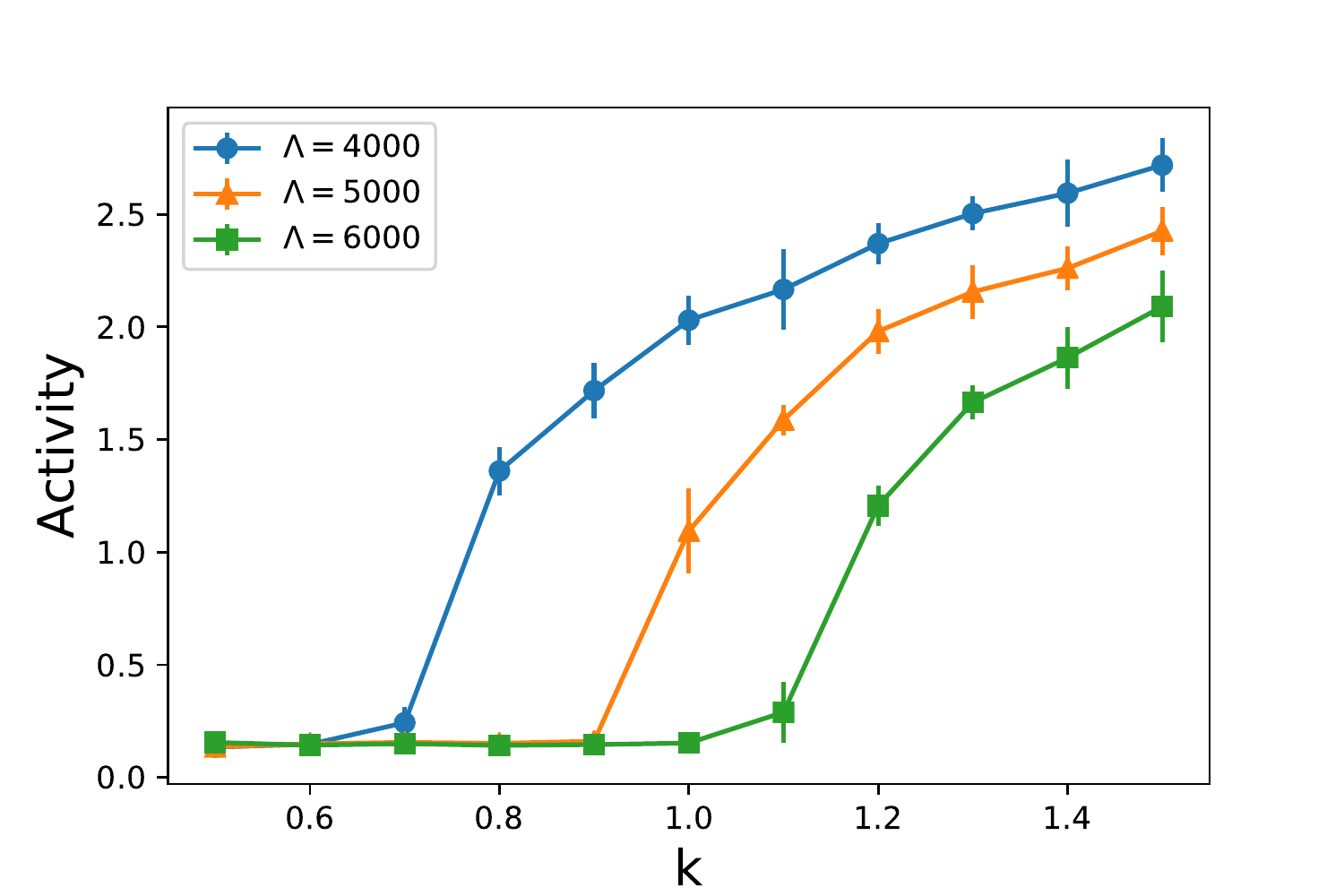}
\caption{Activity of output neuron as a function of $\Lambda$.
(see Fig.~\ref{fig:LineOf3})} \label{fig:compare_lambda}
\end{figure}
It is possible to transmit an impulse through a line several
simulated neurons. Consider the line of three simulated neurons
(Fig.~\ref{fig:LineOf3},~\ref{fig:LINE_3}). In this case we choose
$\varepsilon_1 = k \cdot 1.5 \cdot 10^4, \varepsilon_2 = k \cdot
1.0 \cdot 10^4, \varepsilon_3 = k \cdot 0.5 \cdot 10^4$. As can be
seen from Fig.~\ref{fig:LINE_3},  for small values of the
connection strength $\varepsilon$, the spikes do not pass through
the chain of neurons, but if $\varepsilon$ reaches  some critical
value, the chain becomes transparent for spikes. This effect
allows us to control the transparency of the neural network by the
slight changing of the connection strength $\varepsilon$. Thus, by
controlling the connection strength, we can realize complex
logical connections within our neural network.\par

The dependence of out neuron activity on the parameter $\Lambda$
is shown in Fig.~\ref{fig:compare_lambda}. From this figure it can
be seen that the critical value of the connection strength
$\varepsilon$ depends on the $\Lambda$. Obviously, such dependence
of the critical value of $\varepsilon$ is associated with an
increasing of  kink's action (\ref{kink_act}).\par

\subsection{Logical elements}

We can now turn to the construction of logical elements. In this
subsection we construct from just introduced neurons the simplest
logical elements, such as AND, NOT, OR and XOR. In order to
simplify presentation, we will use the schematic notation for
network elements (Fig.~\ref{fig:schemeSigns}).\par
\begin{figure}
\centering
\includegraphics[width=0.5\linewidth]{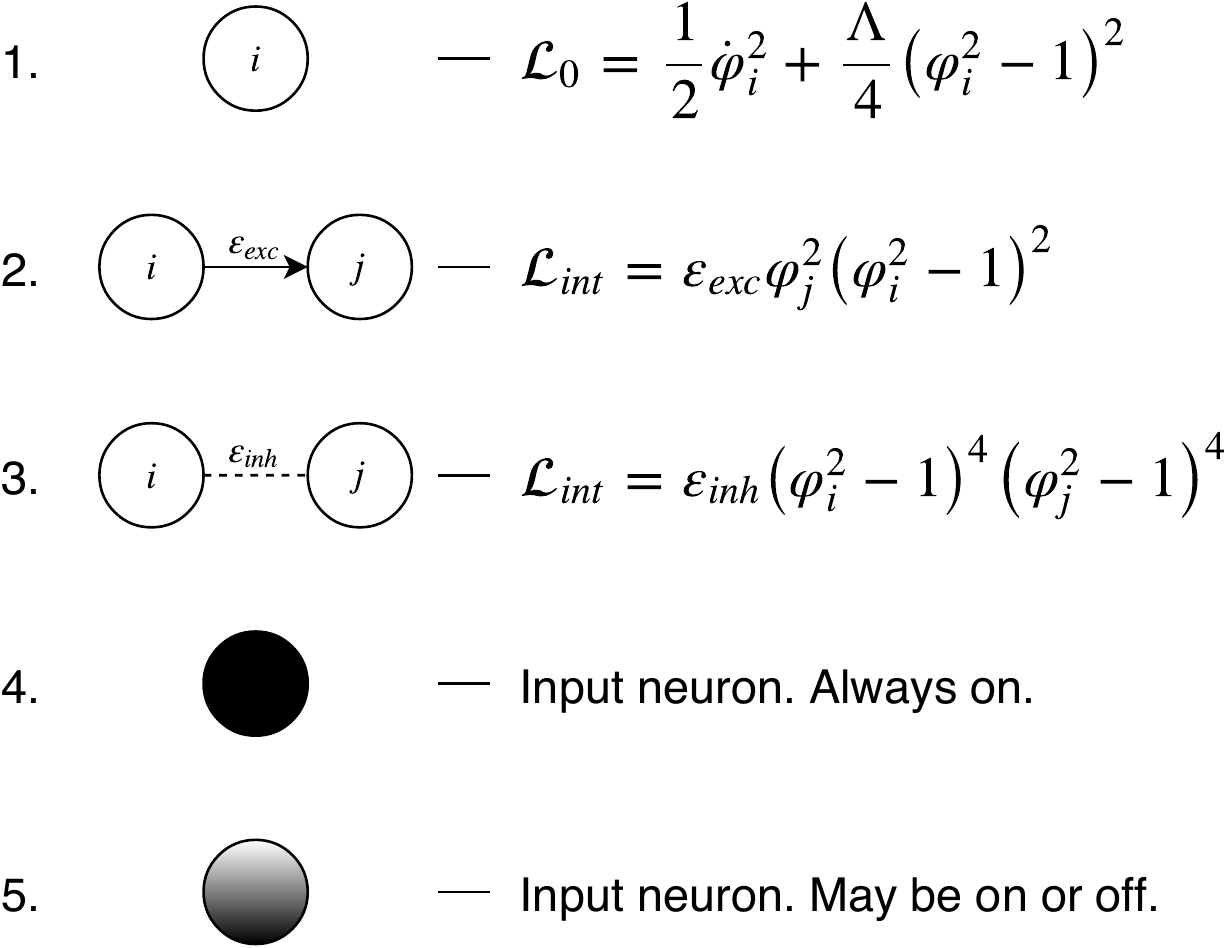}
\caption{Schematic signs used to simplify presentation. 1.
Contribution to the Lagrangian from neuron with index $i$. 2.
Contribution of excitatory connection from neuron $i$ to $j$. 3.
Contribution of inhibiting connection between neuron $i$ to $j$.
4. Input neuron that always active. Its path does not change as
simulation goes. 5. Input neuron that can be either in active or
passive mode. Depending on whether this neuron is active the
network should behave in different way. } \label{fig:schemeSigns}
\end{figure}

\subsubsection{Logical AND}

Logical AND appears to be the simplest element in it's
construction. A neuron connected by a logical AND with some set of
other neurons should experience a spike when all of the neurons it
connected with experience a spike.\par To implement such a
behavior, it is necessary to connect the neuron by excitatory
potential with those neurons whose signals we need to logically
multiply.  It is necessary to choose $\varepsilon_{exc}$
sufficiently small so our output neuron activates only when all of
its inputs are active, and is passive if at least one of input
neuron is passive.\par
\begin{figure}
\centering
\includegraphics[width=0.5\linewidth]{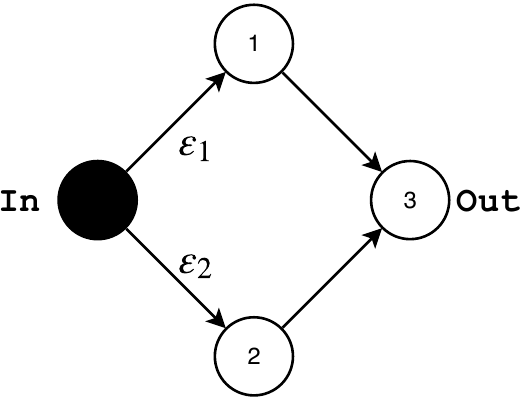}
\caption{Scheme implementing logical AND. Neurons 1 and 2 consume
input and neuron 3 outputs information.} \label{fig:andDemo}
\end{figure}
We demonstrate the operation of such construction in the following
example. In the Fig.~\ref{fig:andDemo} the scheme of such network
is presented. A solid circle indicates the active input neuron.
Circles with numbers are simulated neurons. The arrows show the
connections, and the number next to the arrow corresponds to
$\varepsilon_{exc}$.\par
\begin{figure}
\centering
\includegraphics[width=0.7\linewidth]{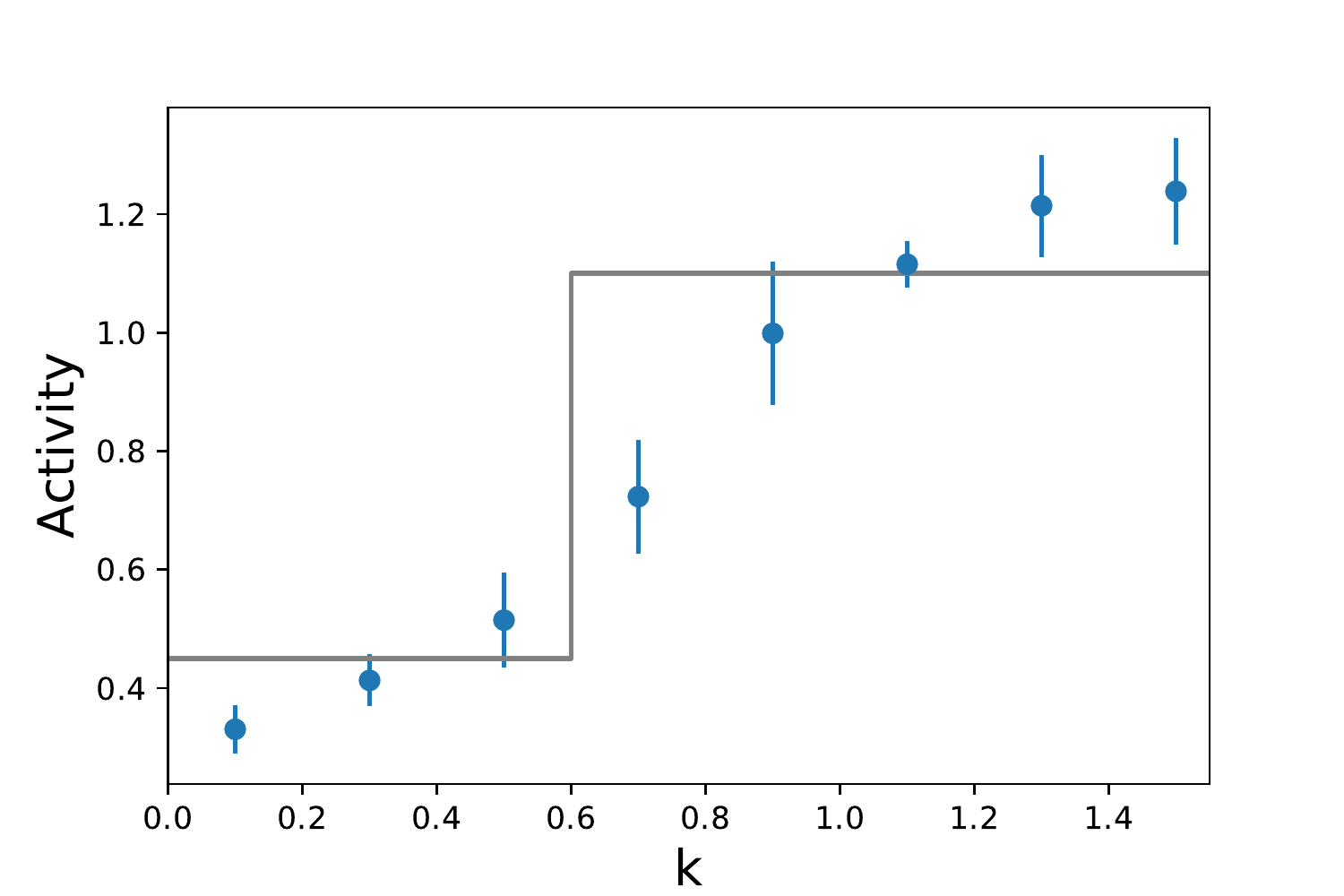}
\caption{The AND case (Fig.~\ref{fig:andDemo}). Activity of output
neuron is nonlinear as function of connection strength scale $k$.
It resembles a smoothed version of step function expected for
discrete case.} \label{fig:andDemoPlot}
\end{figure}
We are interested in two different modes of such a system:
$\varepsilon_1 = \varepsilon_2 = \hat{\varepsilon} = 8000$ (On AND
On should result in On output) and $\varepsilon_1 = 0,
\varepsilon_2 = \hat{\varepsilon}$ (Off AND On should result in
Off output). We choose $\varepsilon_1 = \hat{\varepsilon},
\varepsilon_2 = k\hat{\varepsilon}$. The plot of the activity of
output neuron as function of $k$ is shown in
Fig.~\ref{fig:andDemoPlot}. One may recognize step function in the
output pattern.

\subsubsection{Logical NOT}

Up to this point, we only caused spikes in the neurons, but to
implement arbitrary logic we need to be able to suppress them. For
this purpose, we introduce a logical NOT. We add one more type of
connection, which we call an inhibiting one. Two neurons connected
by such a connection should not spike simultaneously. The
interaction Lagrangian that implements the proposed behavior can
be written as follows:
\begin{equation}
\mathcal{L}_{int} = \varepsilon_{inh}\left(\varphi_{1}^2 - 1\right)^4\left(\varphi_{2}^2 - 1\right)^4.
\end{equation}
\begin{figure}
    \centering
    \begin{minipage}{0.48\textwidth}
        \centering
        \includegraphics[width=\linewidth]{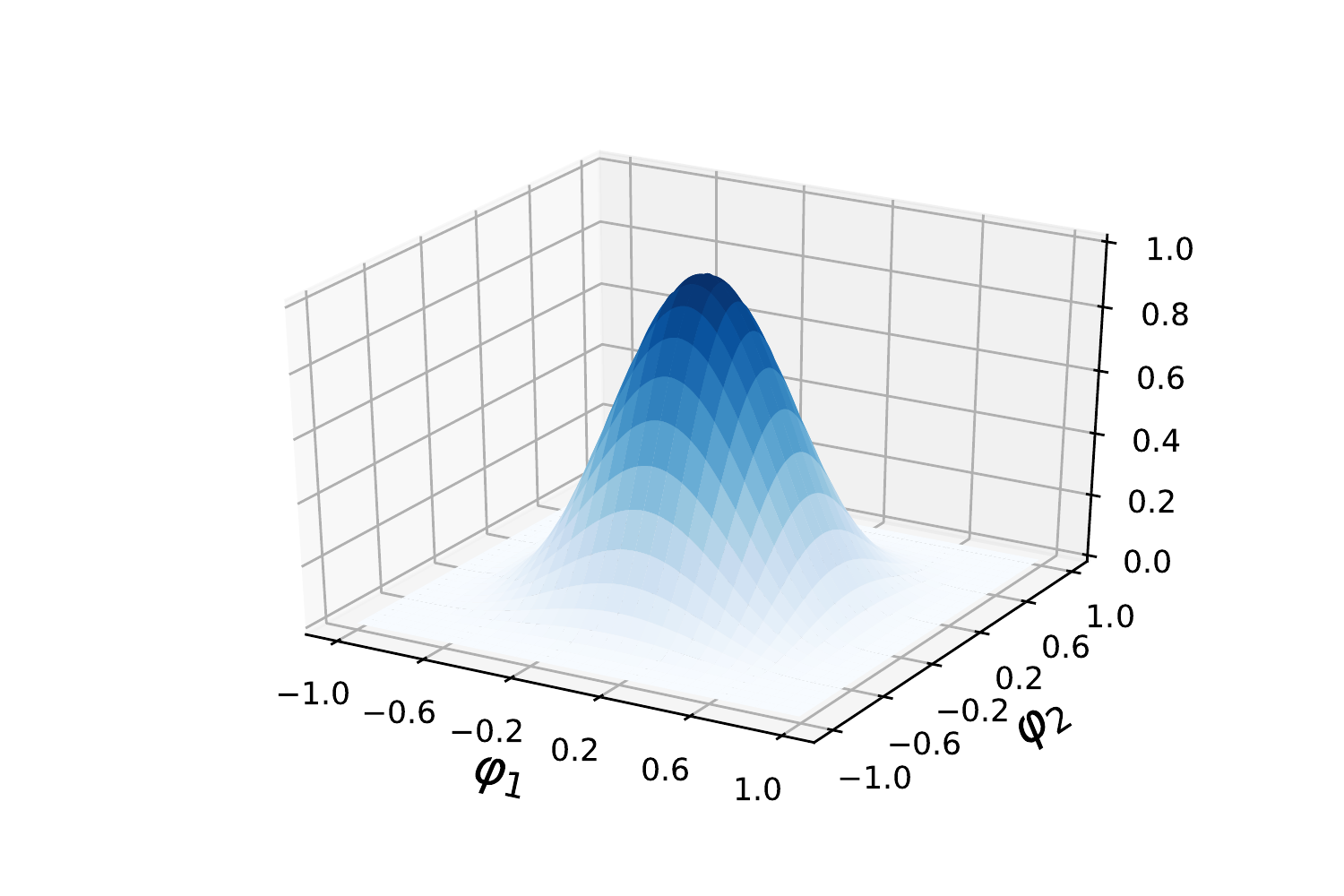} 
        \caption{Inhibiting potential that is used to prevent simultaneous spike of $\varphi_1$ and $\varphi_2$.}
        \label{fig:notPotential}
    \end{minipage}\hfill
    \begin{minipage}{0.48\textwidth}
        \centering
        \includegraphics[width=\linewidth]{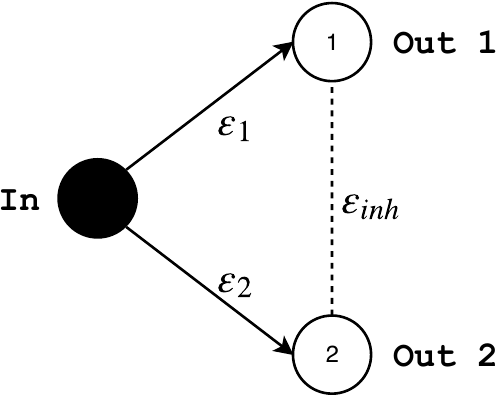} 
        \caption{Scheme implementing logical NOT. Due to $\varepsilon_{inh}$ only
        one of output neurons can be active at the same moment.}
        \label{fig:inhibitingDemoScheme}
    \end{minipage}
\end{figure}

\begin{figure}
\centering
\includegraphics[width=0.7\linewidth]{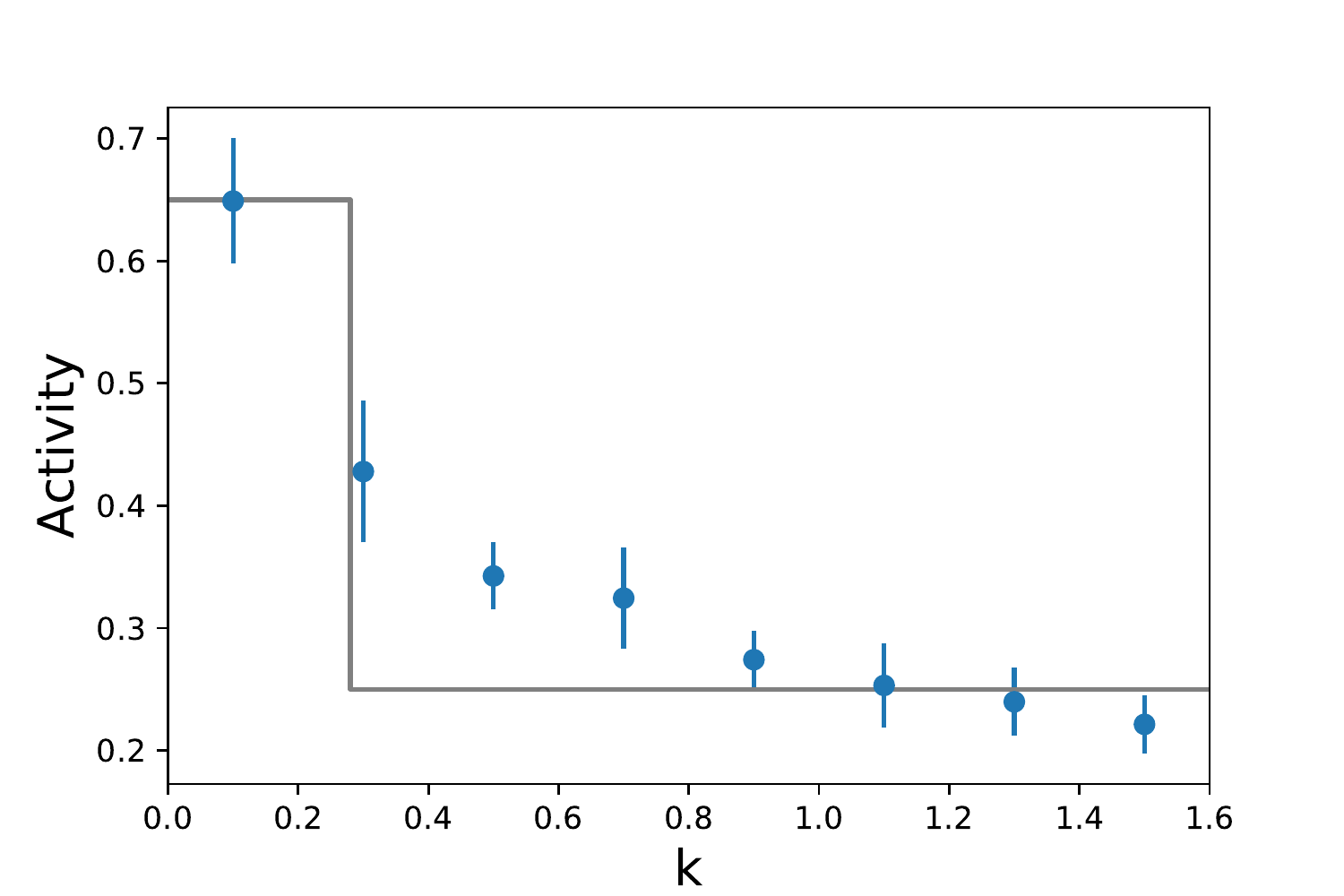}
\caption{The NOT case (Fig.~\ref{fig:inhibitingDemoScheme}). The
activity of output neuron 2. Neuron with weaker connection to
input one is inhibited by another neuron as inhibiting connection
becomes stronger.} \label{fig:logicalNOTexc}
\end{figure}
The plot of the corresponding potential is shown in
Fig.~\ref{fig:notPotential}. Apparently, such interaction has
effect only if both neurons experience a spike.\par To demonstrate
the operation of such a connection, consider the scheme depicted
in Fig.~\ref{fig:inhibitingDemoScheme}. Here the dashed line shows
the inhibiting connection. We set $\varepsilon_1 > \varepsilon_2$.
Activity of neuron 2 as function of $\varepsilon_{inh}$ is
presented in Fig.~\ref{fig:logicalNOTexc}, where
$\varepsilon_{inh} = 50000 k$. As one can see, when
$\varepsilon_{inh} \approx 0$, neuron 2 is active but as
$\varepsilon_{inh}$ grows, neuron 2 becomes inhibited by the
neuron 1.

\subsubsection{Logical OR}

We are now to consider the logical OR. Disjunction in our case
works as follows: an output neuron connected to some set of other
neurons, activates when at least one of the input neurons is
active. At a first glance it may seem that it is enough to simply
connect the neurons with an ordinary exciting connection with
sufficient $\varepsilon_{exc}$ in order to implement such a
behavior. Unfortunately, this can not be done: as we already noted
earlier, appropriate value of $\varepsilon_{exc}$ is restricted
from above which will be violated if all the exciting neurons
experience a spike at the same time (effective $\varepsilon_{exc}$
will be the sum of all the $\varepsilon_{exc}$'s of active
neurons). An appropriate construction is depicted in
Fig.~\ref{fig:orScheme}. It turns out that we should use
intermediate neurons that inhibit each other in such a way that
only one of them can be active at the same time. Only after that
they can be connected to the output neuron. Since the intermediate
neurons become active one by one, the output neuron will never be
overwhelmed.\par
\begin{figure}
    \centering
    \begin{minipage}{0.48\textwidth}
        \centering
        \includegraphics[width=\linewidth]{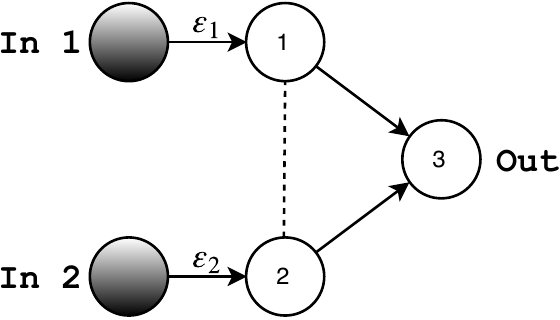} 
        \caption{The logical OR implementation scheme.}
        \label{fig:orScheme}
    \end{minipage}\hfill
    \begin{minipage}{0.48\textwidth}
        \centering
        \includegraphics[width=\linewidth]{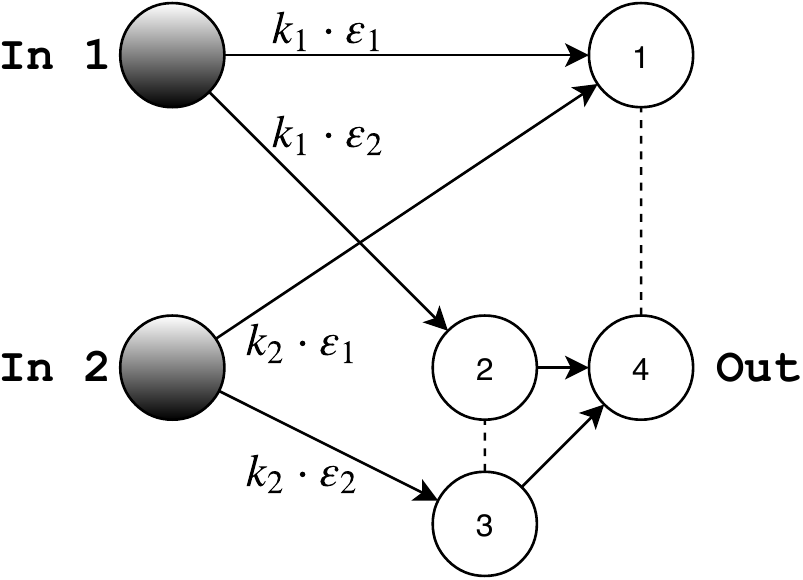} 
        \caption{The logical XOR implementation scheme. $\varepsilon_1 = 3000, \varepsilon_2 = 10000.$}
        \label{fig:xorScheme}
    \end{minipage}
\end{figure}

\begin{figure}
\centering
\includegraphics[width=0.7\linewidth]{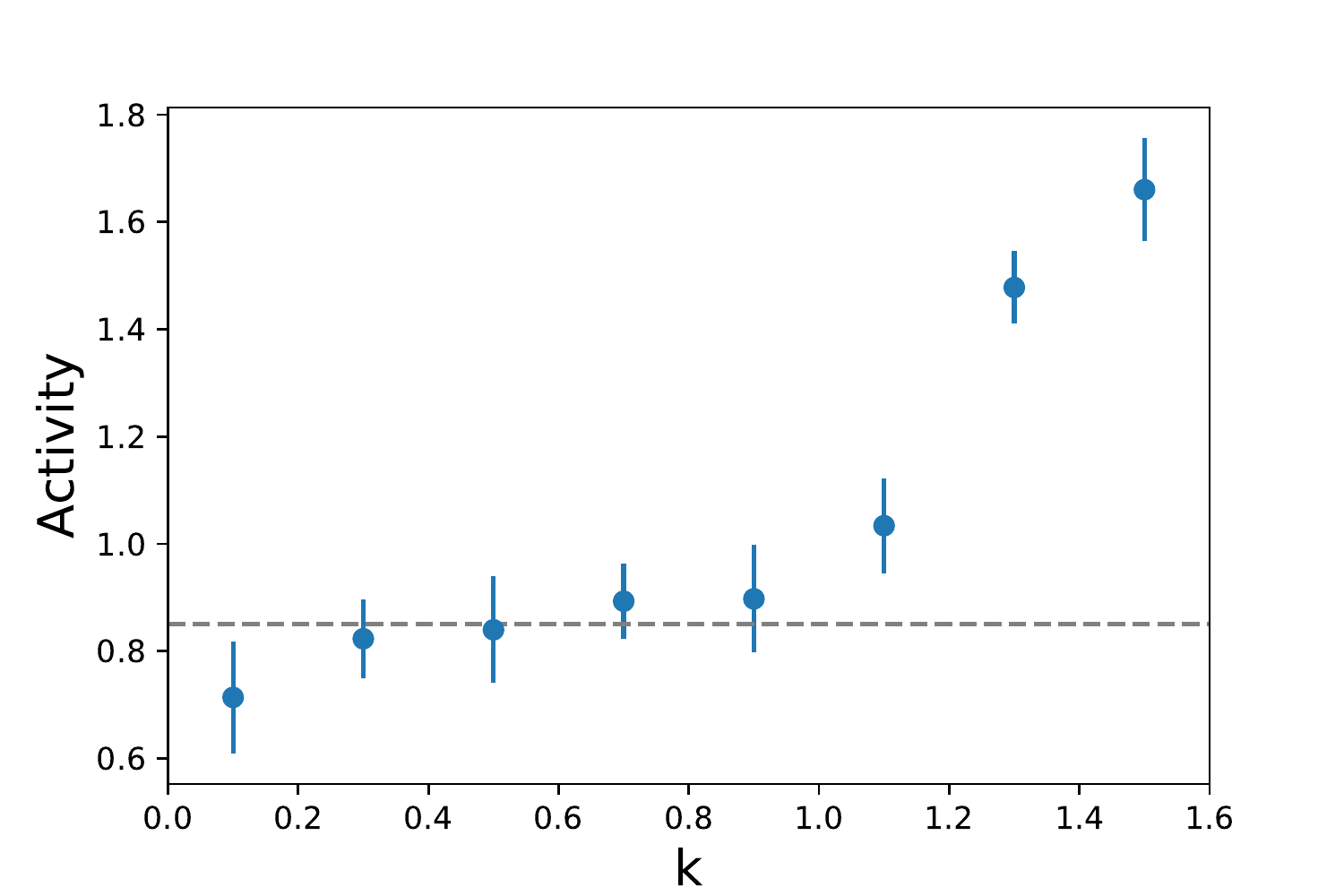}
\caption{The OR case (Fig.~\ref{fig:orScheme}). The activity of
output neuron 3. Output neuron is only affected by strongest input
neuron. When $k=1$ both input neurons are equal.}
\label{fig:logicalOR}
\end{figure}
Again, consider two different modes of operation: $\varepsilon_1 =
\varepsilon_2 = \hat{\varepsilon} = 8000$ (On OR On should result
in On) and $\varepsilon_1 =0, \varepsilon_2 = \hat{\varepsilon}$
(Off OR On should also result in On). We choose $\varepsilon_1 =
\hat{\varepsilon}, \varepsilon_2 = k \hat{\varepsilon}$ and plot
the dependency of the activity of output neuron on $k$
(Fig.~\ref{fig:logicalOR}). Note that one may treat $OR(a, b)$ as
$max(a, b)$: $\forall a, b \in \{0, 1\}: max(a, b) = OR(a, b)$.
And this is what we see on the plot. While $k < 1$ the activity of
output neuron does not change ($activity(max(\hat{\varepsilon},
k\hat{\varepsilon})) \overset{k < 1}{=}
activity(\hat{\varepsilon})$) and as $k$ becomes bigger than 1 the
activity of output neuron also increases.

\subsubsection{An example of construction of logical XOR}

We already have elements from which we can build arbitrary logic,
but before we move on to more complex schemes, we will make sure
that those elements can operate together. To do this, we construct
from them a slightly more complex element - the exclusive or. The
idea of its operation is quite expected: the output neuron should
be active if and only if one of the input neurons is active. \par
We achieve this by means of the scheme depicted in
Fig.~\ref{fig:xorScheme}. Operation of the scheme is very simple:
neuron 4 is active when at least one of the input neurons is
active. But if both are active, neuron 1 will inhibit its
activity. \par
\begin{figure}
\centering
\includegraphics[width=0.7\linewidth]{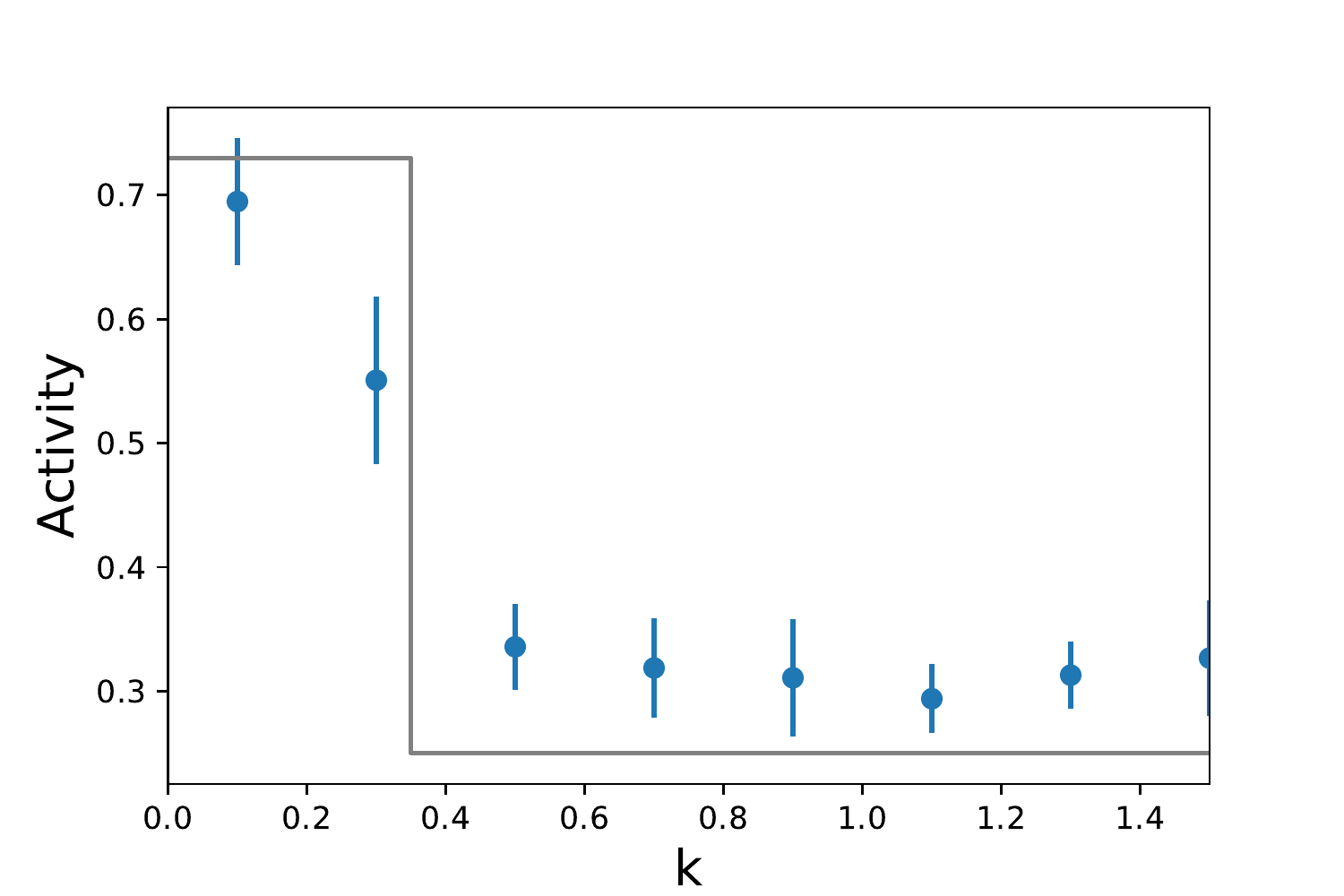}
\caption{The XOR case (Fig.~\ref{fig:xorScheme}).  Activity of
output neuron resembles smoothed step function: it is high if one
of the input neurons is effectively turned off by damping its
connections with small factor of $k$ and low if both neurons are
active.} \label{fig:logicalXOR}
\end{figure}
Let's see how output neuron (4-th) react to change in input's
$\varepsilon$. Activity of output neuron as function of $k_1$ is
shown in Fig.~\ref{fig:logicalXOR}. If first neuron is off ($k_1 =
0$) then activation of second neuron leads to increase of activity
of output neuron. But if first input neuron is active then
activation of second neuron leads to decrease of activity of
output neuron.

\section{Applications of the model}

\subsection{Convolutional model for vertical line detection}

We now move on to a more complex system -- first neural network.
We are to consider convolutional neural network \cite{lecun}
popularized by Krizhevsky et al. \cite{KrizhevskyConv}.\par Basic
operation principles of such a network is as follows: the input is
an image, each pixel depending on its color can be represented by
neuron with varying activity. A convolution operation with some
predetermined kernel is applied to the input image. In our case
the kernel is a matrix of size $3\times3$, its elements can
represent $\varepsilon_{exc}$ or $\varepsilon_{inh}$. We apply
element-wise this matrix to the input image in every possible
position. Depending on the position of the application of the
kernel, the neurons of the input layer will be connected to some
neuron of the second layer.\par The simples application of
convolution model is vertical line detection problem. Let us put
the problem as follows: at the input we have a picture of size
$4\times4$ pixels. The network is supposed to be able to detect a
vertical line in this picture. If there is something else besides
the vertical line on the input image, then the network should not
react to such input. Since the input layer has dimension
$4\times4$ second one should be of size $2\times2$ and third layer
is represented by a single neuron. We connect all neurons of the
second layer to the neuron of third one using $\varepsilon_{exc} =
4000$. To connect the first and second layers we will use the
following kernels:
\begin{equation}
K_{exc} = \begin{pmatrix}
0 & 1 & 0\\
0 & 1 & 0\\
0 & 1 & 0
\end{pmatrix} \times 2000, K_{inh} = \begin{pmatrix}
1 & 0 & 1\\
1 & 0 & 1\\
1 & 0 & 1
\end{pmatrix} \times 15000.
\end{equation}
\begin{figure}
  \center
  \includegraphics[height=0.43\textheight]{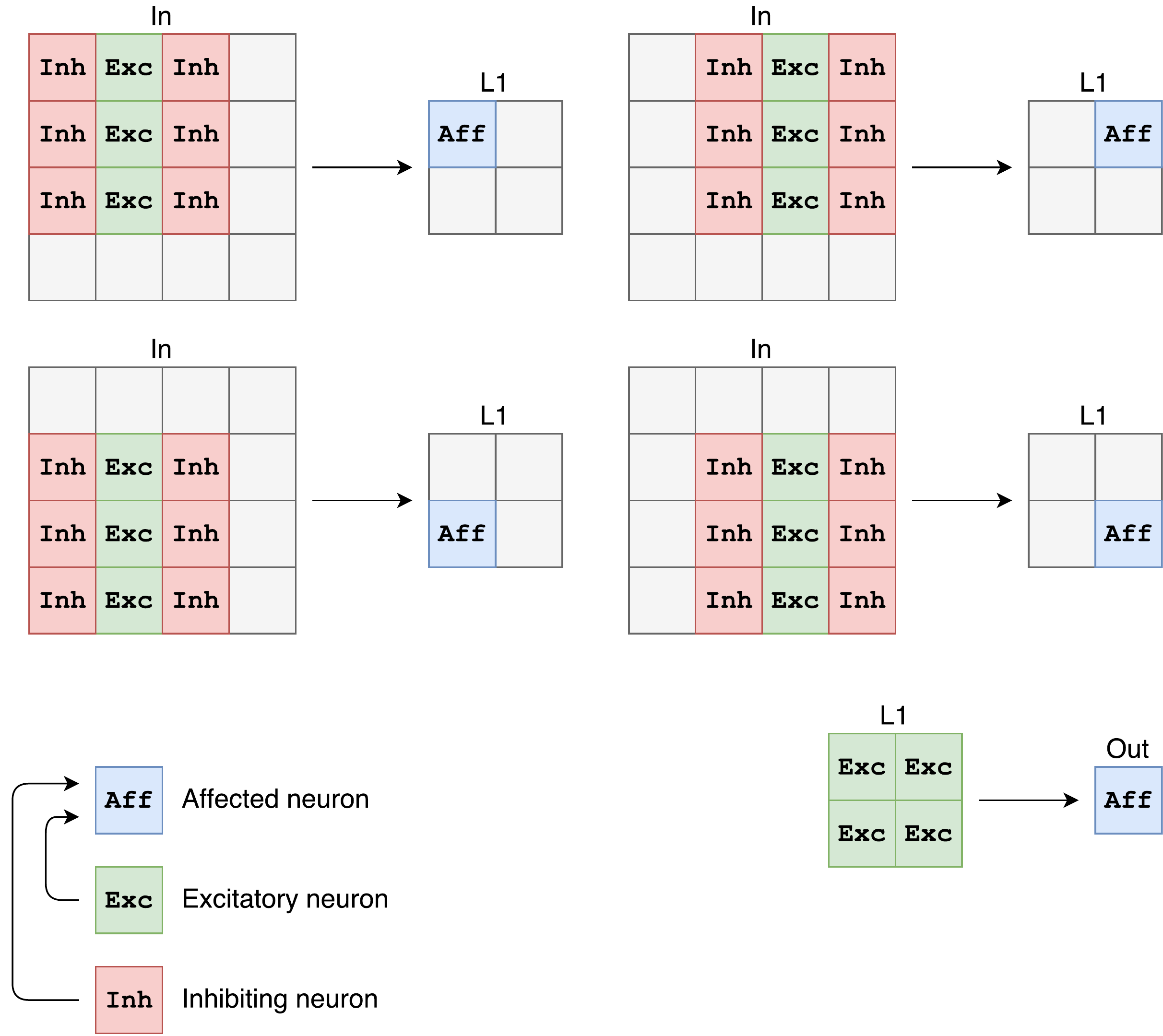}
  \caption{Proposed architecture of a convolutional network.}
  \label{fig:convScheme}

  \center
  \includegraphics[height=0.43\textheight]{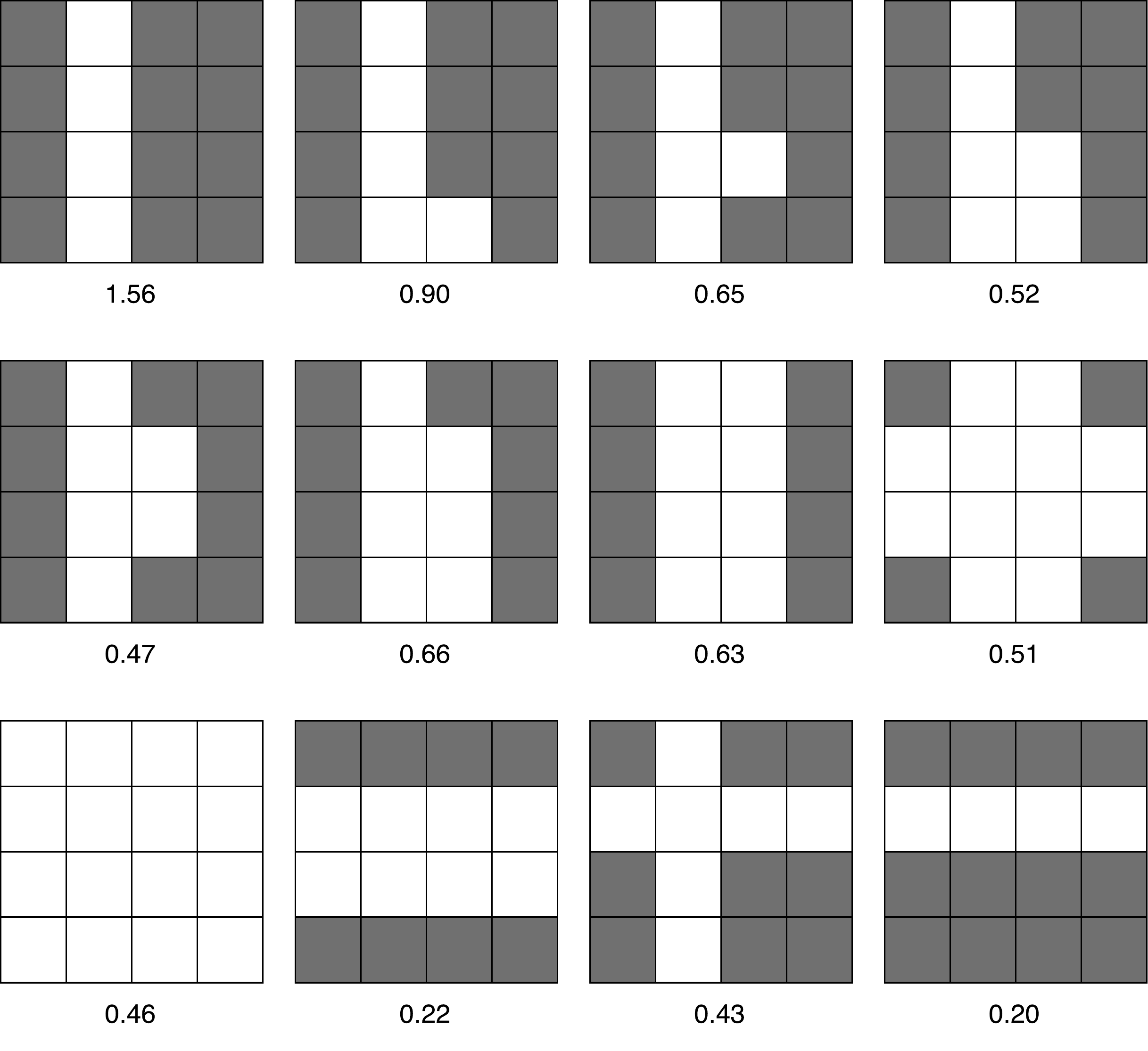}
  \caption{Activity of output neuron is high in case if input resembles vertical line and low otherwise.}
  \label{fig:lineDetection}
\end{figure}
In Fig.~\ref{fig:convScheme} the general scheme of proposed
network, consisting of three layers is depicted, and in
Fig.~\ref{fig:lineDetection} different input images and network
reactions to them are shown. As one can see, the network performs
well and finds a vertical line.\par In real problems, the input
image will have a much larger dimension, which will lead to the
need of increasing the number of layers that will combine the
different kernels, which will represent increasingly complex
images. Simulation of such a network can take a long time, but it
will not have any fundamentally new parts. Thus, we can conclude
that the proposed model can be used for implementing a
convolutional neural networks.

\subsection{Digit recognition}

\begin{figure}[h!]
    \centering
    \begin{minipage}{0.45\textwidth}
        \centering
        \includegraphics[width=\linewidth]{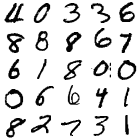}
        \caption{Sample digits from MNIST database.}
        \label{fig:MNISTdigits}
    \end{minipage}\hfill
    \begin{minipage}{0.35\textwidth}
        \centering
        \includegraphics[width=\linewidth]{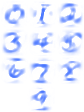}
        \caption{Weights learned by classical neural net for every output (0 to 9).}
        \label{fig:classicWeights}
    \end{minipage}
\end{figure}
The last considered example is the use of our network for the
recognition of handwritten digits. The problem is formulated as
follows: as the input there is an grayscale image of the digit of
the size $28\times28$ pixels, each of which is represented by an
input neuron. At the output, there are 10 neurons, each of which
detects corresponding digit. So our network is the map
$\mathbb{R}^{28\times28} \rightarrow \mathbb{R}^{10}$.\par We we
start with solving this problem using multinomial logistic
regression. Consider $X \in \mathbb{R}^{N \times M}$ where $N$ is
the number of images in training set and $M = 28^2 = 784$ is the
size of an image. So $X_{ij}$ represents the brightness of $j$-th
pixel in $i$-th image. We then map each image into 10 scores
corresponding to numbers using $W \in \mathbb{R}^{M \times 10}$
via matrix multiplication: $S = XW$. So $S_{ij}$ corresponds to
score of $i$-th image treated as $j$-th number. In order to obtain
probabilities we apply softmax function to scores:
\begin{equation}
p_{ij} = \frac{exp(S_{ij})}{\sum_{j=0}^{9} \left( exp(S_{ij}) \right) }.
\label{eq:softmax}
\end{equation}
We can estimate $W$ by minimizing loss function:
\begin{equation}
L = -\frac{1}{N}\sum_{i=0}^{N-1}\sum_{j=0}^{9}ln(p_{ij})\delta^j_{c(i)}
\label{eq:simpleLoss}
\end{equation}
where $c(i)$ is the digit with index $i$ in our dataset. So
$\delta^j_{c(i)}$ is 1 for $j$ corresponding to correct digit and
0 otherwise. We are going to use $W$ as connection strengths
($\varepsilon_{exc}$) in our network so they should not be
negative. So we add term to loss function (\ref{eq:simpleLoss})
penalizing negative weights:
\begin{equation}
L =
-\frac{1}{N}\sum_{i=0}^{N-1}\sum_{j=0}^{9}ln(p_{ij})\delta^j_{c(i)}
+ \lambda \sum_{i=0}^{M-1}\sum_{j=0}^{9}max(-W_{ij}, 0), \lambda
\gg 1.
\label{eq:actualLoss}
\end{equation}\par
To train a classical network, we need many examples of images of
handwritten figures, which should already be properly marked. We
will use the MNIST database\cite{mnistlecun}
(Fig.~\ref{fig:MNISTdigits}). It consists of 60,000 training
samples and 10,000 test samples. Resulting weights $W$ are shown
in Fig.~\ref{fig:classicWeights}. The accuracy of recognition of
the classical network appeared to be $91\%$ on the test set.\par
We can now use $W$ as connection matrix for $\varepsilon_{exc}$
connecting input and output layers of our quantum net. We treat
activity of each neuron as its 'vote' for corresponding number. We
then normalize activities to obtain values that sum to 1 using
softmax function.\par The Lagrangian of the recognition system is
written as follows:
\begin{equation}
\mathcal{L}_{0} =  \sum_{i =
0}^{784}\left[\frac{1}{2}\dot{\hat{\psi}}_{i}^{2} +
\frac{\Lambda}{4} \left(\hat{\psi}_{i}^{2} - 1\right)^2 \right] +
\sum_{j = 0}^{10} \left[\frac{1}{2}\dot{\varphi}_{j}^{2} +
\frac{\Lambda}{4} \left(\varphi_{j}^{2} - 1\right)^2 \right],
\end{equation}
\begin{equation}
\begin{split}
\mathcal{L} = \mathcal{L}_{0} & + \sum_{i = 0}^{784}\sum_{j =
0}^{10} k \hat{\varepsilon}_{ij} \varphi_{j}^{2}
\left(\hat{\psi}_{i}^{2} - 1 \right)^{2}  + \\ & + 10^{-17}
\sum_{k > j}^{10}\sum_{j = 0}^{10} \left(\varphi_{j}^{2} - 1
\right)^{4} \left(\varphi_{k}^{2} -1 \right)^{4},
\end{split}
\end{equation}
where $\hat{\psi}_{i}$ represents the path of neurons
corresponding input image. $\varphi_{j}$ is the coordinates of the
output neurons. $\varepsilon_{ij}$ is the connection strengths of
the input and output layers, we transfer them from the already
trained classical network and normalize to be in $[0, 1]$
according to (\ref{eq:norm}).
\begin{equation}
  \hat{\varepsilon}_{ij} = \frac{W_{ij} - min(W)}{max(W_{ij} -
  min(W))}.
  \label{eq:norm}
\end{equation}
Finally $k=1000$. Note that the first sum in $\mathcal{L}_{0}$ can
be discarded, since it is a constant and does not change in the
modelling process.\par Our input image is grayscale which means
that its parameters may have any value in $[0, 1]$ but up to this
moment we only had binary input neurons (active and passive). We
should somehow incorporate brightness of each pixel into the
parameter of input neuron path: $\hat{\psi_i} = \hat{\psi}(b_i)$
where $b_i$ is the brightness of $i$-th pixel. We choose
$\hat{\psi_i} = \hat{\psi}(b_i) = \sqrt{\sqrt{b_i}\psi^2 -
\sqrt{b_i} + 1}$ where $\psi$ is default path for active neuron
(if $b_i = 0: \hat{\psi_i} = 1$ which is vacuum and if $b_i = 1:
\hat{\psi_i} = \psi$ which is simple active neuron). This makes
possible the next transition (no summation over repeated indices
implied):
\begin{equation}
\hat{\varepsilon}_{ij} \varphi_{j}^{2} \left(\hat{\psi}_{i}^{2} -
1 \right)^{2} = \hat{\varepsilon}_{ij} b_i \varphi_{j}^{2}
\left(\psi_{i}^{2} - 1 \right)^{2} = \varepsilon_{ij}
\varphi_{j}^{2} \left(\psi_{i}^{2} - 1 \right)^{2}
\end{equation}
\par
\begin{figure}
  \centering
  \includegraphics[width=\linewidth]{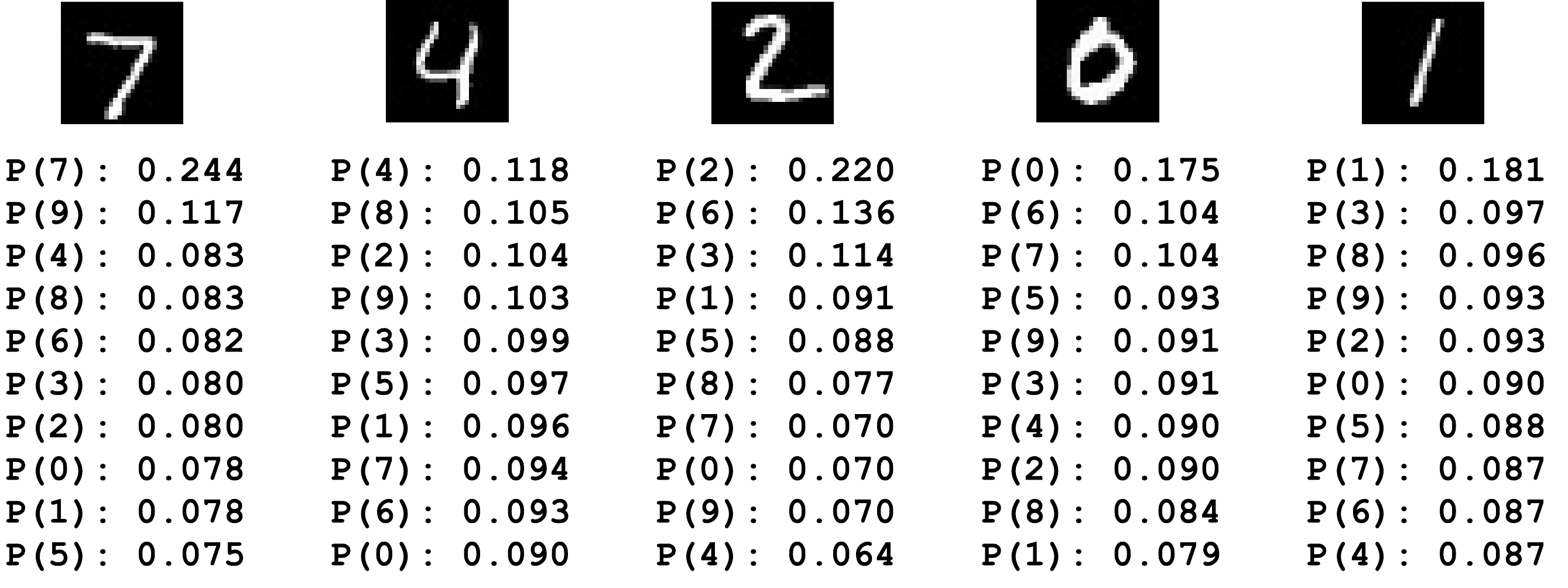}
  \caption{The QM neural network gives higher score to correct digits.}
  \label{fig:quantumDigitRecognitionResults}
\end{figure}
The accuracy of recognition of our quantum-mechanical network
turns out to be somewhat worse than the classical, because in the
process of modelling the Monte Carlo method is used, which adds a
certain portion of randomness. Output of our network is shown in
Fig.~\ref{fig:quantumDigitRecognitionResults}. As can be seen, the
proposed scheme for obtaining $\varepsilon_{exc}$ from the
classical weights is appeared to be satisfactory.

\section{Conclusion}

A neural network  model based on the behavior of a interacting
particles in a two-humped potential was proposed. Excitatory and
inhibitory connections of neurons were introduced. It has been
shown that it is possible to implement the simplest logical
elements using such neurons and their connections. Simplest
convolutional neural network recognizing vertical line has been
constructed. We finally proposed the quantum neural network that
performs digit recognition with satisfactory precession. In the
case of simple architectures the possibility of transferring
weights from classical neural network to a quantum one was
demonstrated.\par
\bibliographystyle{unsrt}
\bibliography{citations}

\begin{thebibliography}{1}

\bibitem{lecun}
Yann LeCun, Bernhard Boser, John~S Denker, Donnie Henderson, Richard~E Howard,
  Wayne Hubbard, and Lawrence~D Jackel.
\newblock Backpropagation applied to handwritten zip code recognition.
\newblock {\em Neural computation}, 1(4):541--551, 1989.

\bibitem{speech1}
A.~Graves, A.~r.~Mohamed, and G.~Hinton.
\newblock Speech recognition with deep recurrent neural networks.
\newblock In {\em 2013 IEEE International Conference on Acoustics, Speech and
  Signal Processing}, pages 6645--6649, May 2013.

\bibitem{google}
Yonghui Wu, Mike Schuster, Zhifeng Chen, Quoc~V Le, Mohammad Norouzi, Wolfgang
  Macherey, Maxim Krikun, Yuan Cao, Qin Gao, Klaus Macherey, et~al.
\newblock Google's neural machine translation system: Bridging the gap between
  human and machine translation.
\newblock {\em arXiv preprint arXiv:1609.08144}, 2016.

\bibitem{polyakov}
Alexander~M Polyakov.
\newblock Gauge fields and strings.
\newblock {\em Contemp. Concepts Phys.}, 3:1--301, 1987.

\bibitem{ceperley}
D.~M. Ceperley.
\newblock Path integrals in the theory of condensed helium.
\newblock {\em Rev. Mod. Phys 67, 279}, 1995.

\bibitem{KrizhevskyConv}
Alex Krizhevsky, Ilya Sutskever, and Geoffrey~E Hinton.
\newblock Imagenet classification with deep convolutional neural networks.
\newblock In F.~Pereira, C.~J.~C. Burges, L.~Bottou, and K.~Q. Weinberger,
  editors, {\em Advances in Neural Information Processing Systems 25}, pages
  1097--1105. Curran Associates, Inc., 2012.

\bibitem{mnistlecun}
Y.~LeCun, L.~Bottou, Y.~Bengio, and P.~Haffner.
\newblock Gradient-based learning applied to document recognition.
\newblock {\em Proceedings of the IEEE}, 86(11):2278--2324, November 1998.

\end{thebibliography}
\end{document}